\documentclass[
 reprint,
superscriptaddress,
 amsmath,amssymb,
 aps,
 pra,
]{revtex4-2}

\usepackage{graphicx}
\usepackage{dcolumn}
\usepackage{bm}
\usepackage[
   colorlinks=true,
   citecolor=blue,
   linkcolor=blue,
   urlcolor=blue,
]{hyperref}

\usepackage{booktabs}%
\usepackage{xcolor}
\usepackage{nicefrac}
\usepackage[utf8]{inputenc}
\usepackage{siunitx}
\usepackage[normalem]{ulem}
\usepackage[final]{changes}
\usepackage[normalem]{ulem}
\newcommand{\hitrap}{HITRAP}
\newcommand{\spectrap}{SPECTRAP}
\newcommand{\TM}{$^{\mathrm{TM}}$}

\newcommand{\operator}[4]{\langle #1 |\boldsymbol{#3}^{(#4)}| #2 \rangle}
\newcommand{\irreducableoperator}[4]{\ensuremath\langle #1 \|\boldsymbol{#3}^{(#4)}\| #2 \rangle}

\def\orcid#1{\kern .08em\href{https://orcid.org/#1}{\includegraphics[keepaspectratio,width=0.7em]{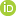}}}

\begin{document}
\preprint{APS/123-QED}

\title[Collinear Laser Spectroscopy of Helium-like $^{11}$B$^{3+}$]{Collinear Laser Spectroscopy of the $1s2s\,^3\!S_1\rightarrow 1s2p\,^3\!P_{0,2}$ Transitions in Helium-like $^{11}$B$^{3+}$}
\author{Axel~Buß}
\thanks{These two authors contributed equally.}
\affiliation{Institut für Kernphysik, Universität Münster, 48149 Münster, Germany}
\author{Konstantin~Mohr\orcid{0000-0002-2669-6857}\textsuperscript{*}}
\email{k.mohr@gsi.de}
\affiliation{Institut f\"ur Kernphysik, Technische Universit\"at Darmstadt, 64289 Darmstadt, Germany}
\affiliation{Helmholtz Research Academy Hesse for FAIR, GSI Helmholtzzentrum für Schwerionenforschung GmbH, 64291 Darmstadt, Germany}
\author{Volker~Hannen\orcid{0000-0002-2944-8373}}\email{hannen@uni-muenster.de}
\affiliation{Institut für Kernphysik, Universität Münster, 48149 Münster, Germany}
\author{Zoran~Andelkovic\orcid{0009-0003-0576-4317}}
\affiliation{GSI Helmholtzzentrum für Schwerionenforschung GmbH, 64291 Darmstadt, Germany}
\author{Max~Horst\orcid{0000-0002-7753-7557}}
\affiliation{Institut f\"ur Kernphysik, Technische Universit\"at Darmstadt, 64289 Darmstadt, Germany}
\affiliation{Helmholtz Research Academy Hesse for FAIR, GSI Helmholtzzentrum für Schwerionenforschung GmbH, 64291 Darmstadt, Germany}
\author{Phillip~Imgram\orcid{0000-0002-3559-7092}}
\affiliation{Institut f\"ur Kernphysik, Technische Universit\"at Darmstadt, 64289 Darmstadt, Germany}
\author{Kristian~König\orcid{0000-0001-9415-3208}}
\affiliation{Institut f\"ur Kernphysik, Technische Universit\"at Darmstadt, 64289 Darmstadt, Germany}
\affiliation{Helmholtz Research Academy Hesse for FAIR, GSI Helmholtzzentrum für Schwerionenforschung GmbH, 64291 Darmstadt, Germany}
\author{Bernhard~Maass\orcid{0000-0002-6844-5706}}
\affiliation{Institut f\"ur Kernphysik, Technische Universit\"at Darmstadt, 64289 Darmstadt, Germany}
\affiliation{Argonne National Laboratory, Lemont, IL 60439, USA}
\author{W.~Nörtershäuser\orcid{0000-0001-7432-3687}}
\affiliation{Institut f\"ur Kernphysik, Technische Universit\"at Darmstadt, 64289 Darmstadt, Germany}
\affiliation{Helmholtz Research Academy Hesse for FAIR, GSI Helmholtzzentrum für Schwerionenforschung GmbH, 64291 Darmstadt, Germany}
\author{Simon~Rausch\orcid{0000-0002-8833-0121}}
\altaffiliation{Present address: Department of Physics,  University of Jyväskylä, 40014 Jyväskylä, Finland}
\affiliation{Institut f\"ur Kernphysik, Technische Universit\"at Darmstadt, 64289 Darmstadt, Germany}
\affiliation{Helmholtz Research Academy Hesse for FAIR, GSI Helmholtzzentrum für Schwerionenforschung GmbH, 64291 Darmstadt, Germany}
\author{Rodolfo~Sánchez\orcid{0000-0002-4892-4056}}
\affiliation{GSI Helmholtzzentrum für Schwerionenforschung GmbH, 64291 Darmstadt, Germany}
\author{Christian~Weinheimer\orcid{0000-0002-4083-9068}}
\affiliation{Institut für Kernphysik, Universität Münster, 48149 Münster, Germany}
\date{\today}

\begin{abstract}
We report on hyperfine structure measurements of the $1s2s\,{}^{3\!}S_1 \rightarrow 1s2p\,{}^{3\!}P_{0,2}$ fine-structure transitions in the spectrum of  $^{11}$B$^{3+}$ ions. The helium-like ions were created in an electron beam ion trap (EBIT) and ejected after a short production period of 15\,ms. A challenge of the experiment was the treatment of the complex energy-time-profile of the created ion bunches and the sensitivity of the collinear laser spectroscopy to the starting potential inside the EBIT. Despite these complications and the low statistics of the experiment, we were able to confirm and improve the results of a previous measurement of these transitions. While we find moderate tension in the results for the hyperfine parameters of the $^{3\!}P_{0,2}$ levels, their fine-structure level energies are in reasonable agreement between experiments and with theory. 
\end{abstract}

\maketitle

\section{Introduction}
\label{sec:intro}
The two-electron systems of helium and He-like ions are of interest in atomic and nuclear physics. Calculations in hydrogen-like systems can be performed with extraordinary accuracy but nuclear structure sets the ultimate floor and already dominates in hyperfine structure and muonic systems~\cite{Udem.1997,Shabaev.2001,Eides.2001,Karshenboim.2005,Volotka.2005,Volotka.2014b,Ullmann.2017,Beyer.2017,Udem.2018,Yerokhin.2019,Antognini.2022,Pachucki.2024}. Adding only one more electron complicates ab initio calculations considerably~\cite{Drake1988,Drake.1999,Drake.2002,Pachucki.2009,Pachucki2010,Yerokhin.2010,Yerokhin.2018,Yerokhin.2022,Drake.2026}. One reason is the existence of analytical solutions of the Dirac equation for the two-body hydrogen-like problem which can be appended by quantum electrodynamical (QED) and nuclear structure corrections. 
This is not the case anymore for the three bodies in a He-like system. Here, one has to start with numerical solutions, and the calculation of electron-electron correlations is notoriously difficult. However, impressive progress has been made in recent years. For helium, calculations have been advanced so far that the charge radius is in principle extractable from frequencies measured in $1s2s\rightarrow 1s2p$ transitions~\cite{Yerokhin.2010,Yerokhin.2018,Yerokhin.2022} but inconsistencies between different transitions in experiment and theory still hamper a vital test of the results in comparison to charge radii obtained in elastic electron scattering and muonic atoms~\cite{Patkos.2021}. 

With increasing $Z$, the relative contribution of the uncertainty from electron-electron correlations increases and complicates a similar approach as it has been recently demonstrated in He-like carbon~\cite{Imgram2023}. The charge radius extracted for $^{12}$C is in good agreement with the determinations from elastic scattering and muonic atoms, but has a too large uncertainty to resolve the small discrepancy between the radii extracted using these two approaches. The latter would become possible if theoretical accuracy is improved by two orders of magnitude, which requires at least calculating the most important $m\alpha^8$ contributions. This is different in the case of boron, where the charge radius is only known with a relative accuracy on the percent level compared to 0.01\% in $^{12}$C~\cite{Fricke.2004}. 

Several trials are currently ongoing to improve the situation for the boron isotopes. The QUARTET collaboration is preparing new muonic-atom measurements~\cite{Ohayon.2024,Eizenberg2026} to establish absolute radii with higher precision. We are working towards high-precision measurements on the full manifold of $1s2s\,^3\!S_1 \rightarrow 1s2p\,^3\!P_J$  transitions in He-like boron $^{10,11}$B$^{3+}$ to extract absolute charge radii, but also differences in the mean-square charge radii with even higher precision. Towards this goal, we present first \hbox{(anti-)}collinear measurements of the hyperfine structure in the $1s2s\,^3\!S_1 \rightarrow 1s2p\,^3\!P_{0,2}$ transitions of $^{11}$B$^{3+}$ ions produced in an electron beam ion trap (EBIT). 

These first experiments demonstrated the production of metastable B$^{3+}$ ions using the Metal Ion from Volatile Compounds (MIVOC) technique with trimethylborate $\mathrm{(CH_3O)_3B}$ inside an EBIT and the sufficient population of the $1s2s\,^{3\!}S_1$ level for laser spectroscopy. 
The measurements were performed at the \hitrap\ platform at the GSI Helmholtz Centre for Heavy Ion Research in Darmstadt, Germany. This platform supports a number of  experiments that will obtain highly charged heavy ions up to bare or hydrogen-like uranium produced by the GSI accelerator facility, decelerated in the ESR and the \hitrap\ facility, finally electron-cooled in the \hitrap\ Cooling Trap before delivered to the experimental platform~\cite{Kluge.2008,Herfurth2015,Rausch.2026}. For commissioning of these experiments a commercial EBIT (SPARC-EBIT)~\cite{Sokolov2010,ORourke2009} is installed on the platform and has been used to produce the He-like boron ions. The SPECTRAP experiment~\cite{Andelkovic2012,Schmidt.2018b}, ultimately designed to perform laser spectroscopy on H- and Li-like bismuth $^{209}$Bi$^{82+,80+}$ ions in a Penning trap, has been modified to carry out collinear laser spectroscopy on ions delivered from the EBIT by implementing the prototype of the new optical detection chamber built for the CRYRING@ESR storage ring~\cite{Lestinsky.2016}. Even though we had to analyze a complex lineshape and obtained relatively low statistics in the experiment, we were able to slightly improve the uncertainty of previously measured transition frequencies of several hyperfine components in the targeted transitions.
\section{Experimental Setup}
\subsection{Ion Beam Production and Transport}
\label{ssec:ebit}
%
He-like boron ions ($^{10,11}$B$^{3+}$) were produced in a Dresden EBIT from DREEBIT GmbH~\cite{dreebit}. Previously used for Ar$^{16+}$~\cite{Rausch2022} and highly charged potassium ions~\cite{Sokolov2010}, the EBIT was operated here using the MIVOC method, which introduces volatile metal compounds. Trimethyl borate (vapor pressure $\approx 200$\,mbar at $25^{\circ}\mathrm{C}$) was supplied via a needle valve, maintaining a stable pressure of $2\times10^{-9}$\,mbar regulated by a PID loop.
Inside the EBIT, molecules are dissociated and ionized by the electron beam. The ions are confined radially by the beam’s space charge and axially by electrostatic barriers. The central trap potential of $+4\,$kV largely determines the extraction energy. During charge breeding, the ion charge state increases until the ionization energy exceeds the electron energy and/or an equilibrium between ionization and recombination reactions is reached. For $^{11}$B$^{3+}$, a breeding time of 15\,ms was optimal
and the extracted ion bunches were $<$\SI{2}{\micro\second} wide.

Transport to the SPECTRAP experiment was achieved via a $\sim 13$-m-long beamline~\cite{Andelkovic2015}. Mass-to-charge selection was performed with a magnetic multi-passage spectrometer that was operated as a $90^{\circ}$ sector magnet to inject the bunches into the electrostatic beamline and subsequent time-of-flight diagnostics. Quadrupole doublets refocused the beam, while two electrostatic kicker benders deflected it by $90^{\circ}$ each, the second guiding the ions vertically into the detection chamber, where the ion beam is overlapped with a laser beam.

\subsection{Detection System}
\begin{figure}[h]
	\centering
	\includegraphics[width=0.9\columnwidth]{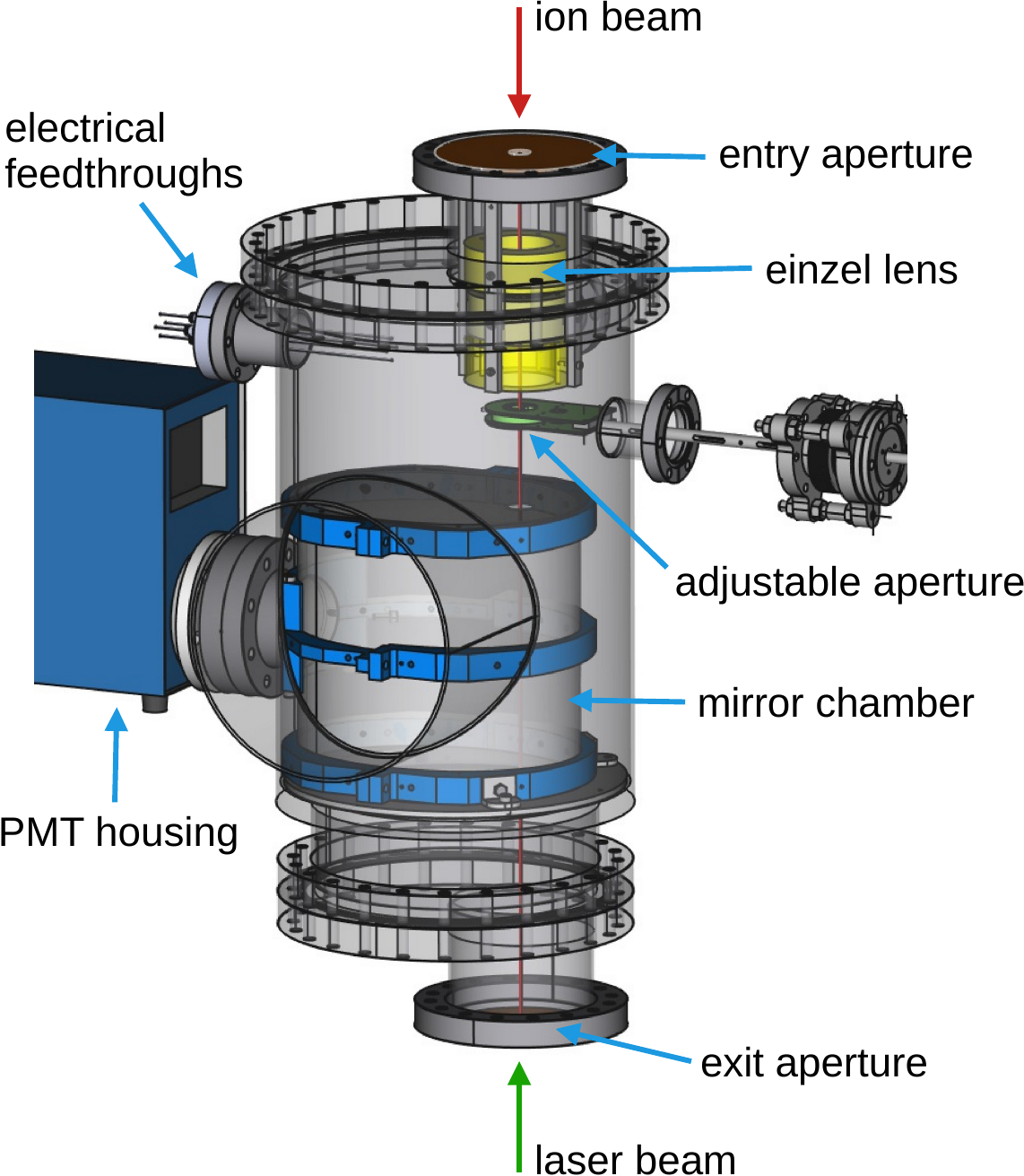}
	\caption{Fluorescence detection chamber. The counter-propagating ion and laser beams enter the chamber via their respective apertures. The ion beam can be manipulated using a segmented einzel lens and an adjustable aperture above the mirror chamber. Fluorescence photons emitted by the ions are collected by a mirror system and focused onto a PMT outside the vacuum. The mirror chamber can be put onto an electric potential to accelerate or decelerate the ions before interaction with the laser beam. Using a viewport on top of the beamline, a mirror can be installed to reflect the laser light back into the mirror chamber alongside the ion beam in order to perform measurements in a collinear geometry.}
	\label{fig:spectrap_detector}
\end{figure}

The fluorescence detection region (FDR) being used is shown in Fig.\,\ref{fig:spectrap_detector} and is the prototype of a similar fluorescence detection system developed for laser spectroscopy measurements at the CRYRING@ESR storage ring~\cite{Mohr2025}. 
It was installed between the cryostat of the \spectrap\ experiment and the last electrostatic bender of the connecting beamline. Its main components are an elliptical mirror inside a vacuum chamber and a UV-sensitive photomultiplier tube (PMT) of type 9235QA by \textit{ET Enterprises Ltd.} in a housing outside the vacuum (see Fig.\,\ref{fig:spectrap_detector}).
The mirror chamber has \SI{20}{mm} diameter openings in its end-caps for passage of laser and ion beam and is electrically isolated from the vacuum chamber. This allows to apply electrostatic potentials up to \SI{\pm 500}{V} in order to accelerate or decelerate the ions before entering the mirror chamber to scan the laser frequency by the corresponding Doppler shift in the rest frame of the ion (Doppler tuning) as discussed in Sec.~\ref{ssec:dshift_scan}. 
To be able to perform measurements also in collinear geometry, a mirror was temporarily installed above a viewport on top of the beamline to reflect the laser light back into the detection chamber, now copropagating with the ion beam. This, however, required an extra careful adjustment of the optical components with the laser beam now passing the setup twice on its way up and down again. This was used to determine the transition frequency of the strongest hyperfine structure component, which provided then a reference line to determine the ion energy as described below.\\
The dominant source of background signal was laser stray light within the mirror chamber, largely originating from reflections of the laser beam at the entrance and exit window surfaces. The fixed-size copper apertures at the top and bottom of the setup were most effective in reducing this background, while the adjustable aperture between the last einzel lens and the mirror chamber had only little effect but was an essential tool for setting up the beam overlap.
\subsection{Laser System}
\begin{figure}[tbp]
 \centering
 \includegraphics[width=\columnwidth]{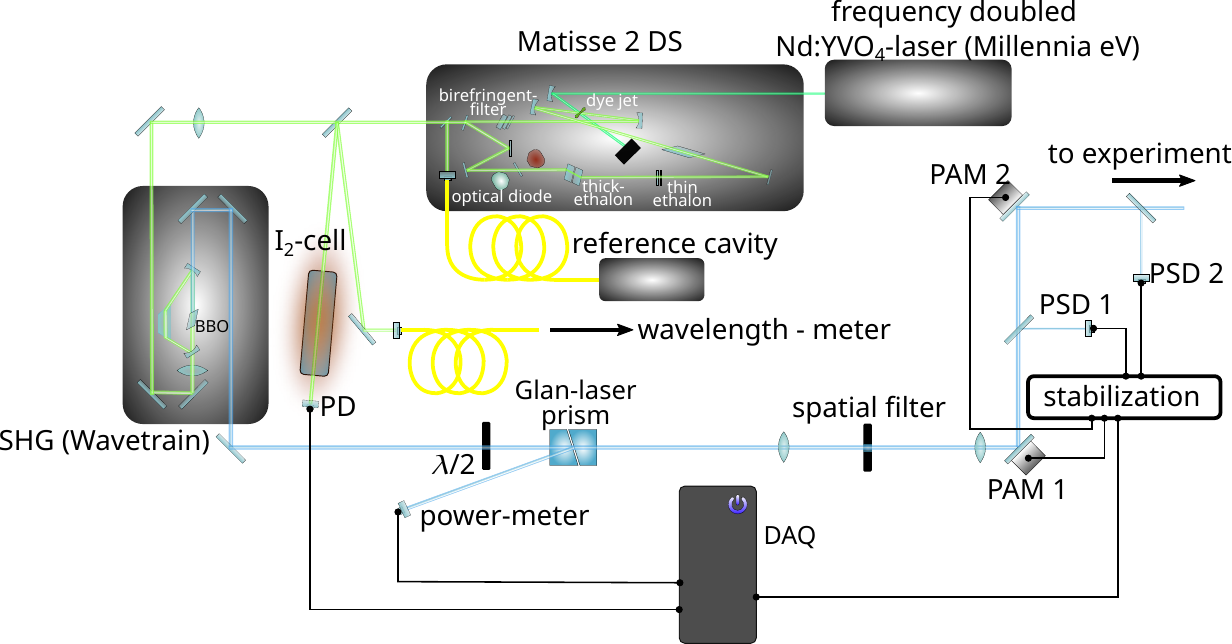}
 \caption[Laser system schematic]{Schematic view of the laser system taken from~\cite{Mohr2022}. For details see text. SHG: second harmonic generation, $\lambda/2$: waveplate, PSD: position sensitive detector, PAM: piezo actuated mirror,  PD: photodiode.
 }
 \label{fig:Lasersystem-pdf}
\end{figure}

The laser system is schematically shown in Fig.\,\ref{fig:Lasersystem-pdf} 
and described in more detail in~\cite{Mohr2022}. 
A ring dye-laser model Matisse 2 DS (SIRAH Lasers) operated with \SI{1.93}{g/l} Rhodamin 110 dye dissolved in ethylen                                             glycole is pumped by \SI{8}{W} of a Millennia eV20 Nd:YVO$_4$ pump laser and produces about \SI{0.8}{W} at a wavelength of about 564 nm. A small fraction of its output is used for active long-term frequency stabilization to a High-Finesse\TM WSU-10 wavelength meter.
The largest part of the laser power is fed into the enhancement cavity of a Wavetrain\TM\ for second-harmonic generation inside a $\beta$-barium-borate (BBO) crystal to produce the required (Doppler-shifted) wavelengths for the targeted B$^{3+}$ transitions at about \SI{282}{nm}. 
A spatial filter is used to remove halos and higher-order spatial modes that might generate stray light in the optical detection region. The commercial beam stabilization system from MRC Systems GmbH stabilized the laser beam in angle and position. The laser power is monitored with a power meter and continuously recorded with the data acquisition system.
\section{Data Taking} 
\label{ssec:dshift_scan}

Doppler tuning is the generally preferred method for recording resonances in collinear laser spectroscopy. With ion and laser beams aligned either co-propagating ($\theta = 0$) or counter-propagating ($\theta = \pi$), a change in the ion velocity $\beta$ converts the laboratory laser frequency $\nu_{\rm lab}$ into the rest-frame frequency $\nu_0$ via
\begin{equation}
    \nu_0 = \nu_{\rm lab}\,\gamma\,(1-\beta\cos\theta),
    \label{eq:Doppler-Shift}
\end{equation}
where $\gamma = 1/\sqrt{1-\beta^2}$. The velocity follows from the acceleration voltage $U$ as
\begin{equation}
    \beta = \sqrt{1 - \left(1 + \frac{qU}{mc^2}\right)^{-2}},
\end{equation}
and a small voltage change $\Delta U$ induces an approximate Doppler shift
\begin{equation}
    \Delta\nu_{\rm lab} \approx \pm \frac{q\nu_0}{\sqrt{2qUmc^2}}\,\Delta U,
    \label{eq:DifferentialDopplerShift}
\end{equation}
with the sign depending on collinear ($+$) or anticollinear ($-$) geometry. For the parameters of this experiment ($U\approx 4\,$kV, $q=3e$, $m\approx 11$\,amu, $\nu_0\approx 1062$\,THz), this yields $\Delta\nu_{\rm lab}/\Delta U \approx \pm 203\,\text{MHz/V}$.
Doppler tuning offers a large scan range using modest voltages, allows the laser to remain frequency-stabilized, and enables rapid skipping of wide frequency intervals—useful for large hyperfine splittings. However, at low beam energies, large tuning voltages can perturb beam steering and thus the ion–laser overlap. Since such effects were observed at high tuning voltages, Doppler tuning was used only for locating the hyperfine features and optimizing EBIT parameters. Once suitable conditions were found, the measurements were switched to laser-frequency scanning with a constant, low acceleration voltage in the fluorescence region, preventing resonant interaction and optical pumping outside the detection chamber.

A high-statistic measurement was carried out on the $^{3\!}S_{1}\, (F=\nicefrac{5}{2}) \rightarrow {^3\!}P_{2}\, (F=\nicefrac{7}{2})$ hyperfine transition in $^{11}$B$^{3+}$, which is the strongest hyperfine transition in this manifold. It provides a set of reference data for determining analysis parameters for this as well as for all other transitions measured with lower statistics. The result is shown in the upper panel of Fig.\,\ref{fig:overnight_contour}a where the signal rate is recorded as a function of the arrival time of the ions at the detector ($y$-axis) and the fundamental frequency of the dye laser ($x$-axis). The width of the recorded arrival-time window was set to \SI{10}{\micro\second} of which only a \SI{1.6}{\micro\second} long slice is shown in the figure, excluding regions where the observed rate is purely caused by laser background. 
The two-dimensional shape of the resonance signal exhibits a central maximum and two distinct tails extending towards higher frequencies. These tails originate from slower ions, which are resonantly excited at higher laser frequencies due to their lower Doppler shift. The nearly horizontal tail is attributed to the switching process and quantitatively follows the relation between the expected arrival time and the shift in the resonance frequency.
The second tail, characterized by a steeper slope, can be explained by ions with lower thermal energy remaining longer in the EBIS reservoir prior to extraction. In addition, energy loss due to inelastic scattering with residual gas atoms along the beam path may further contribute to this tail.

\section{Quantitative Description of the Spectra}
\label{ch:B_analysis}
\begin{figure}[htp]
 \centering
 \includegraphics[width=\columnwidth]{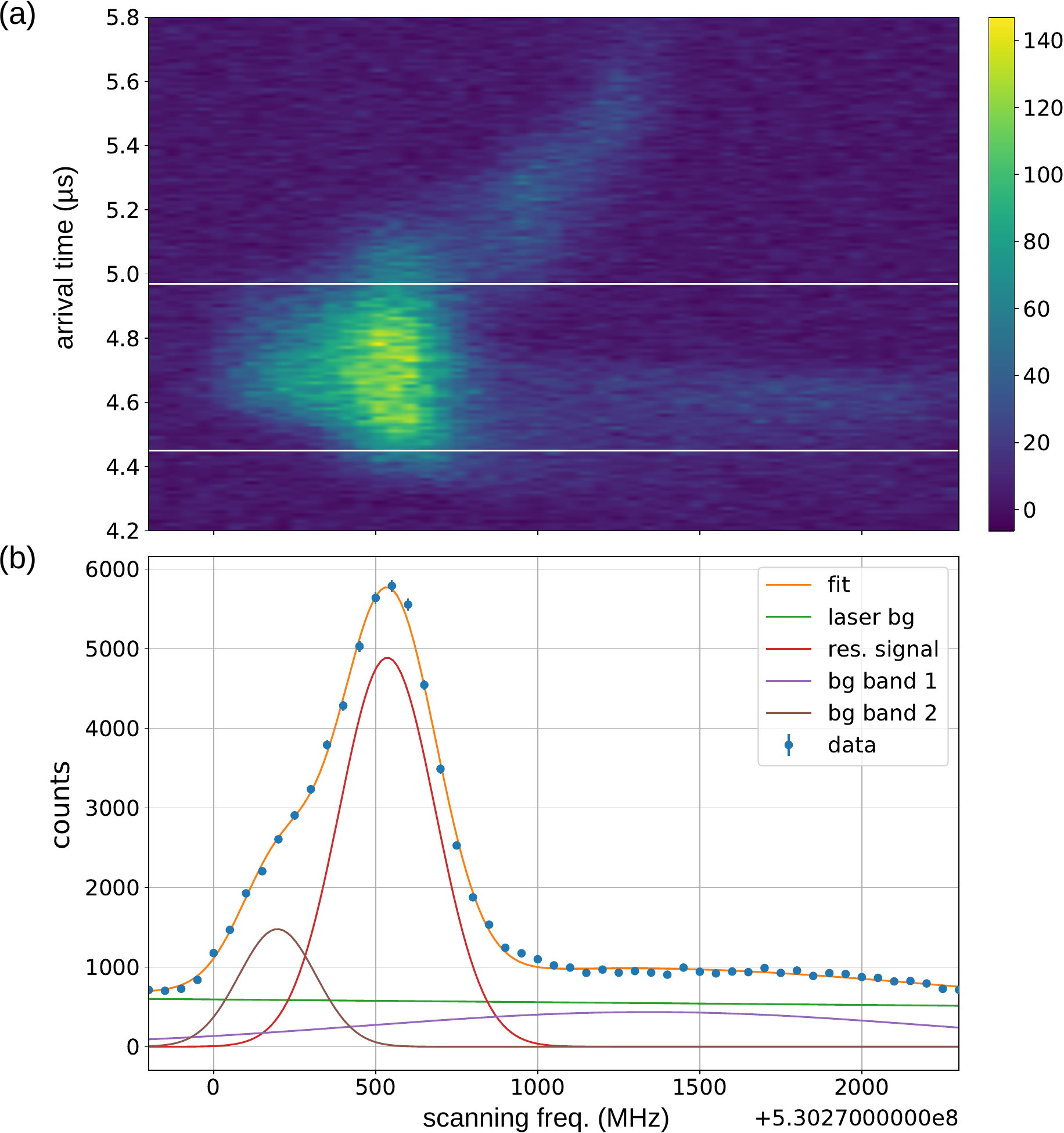}
 \caption{(a) Histogram of the arrival-time distribution of the fluorescence photons as function of the scanning frequency for the dominant $^{3\!}S_{1}\, (F=\nicefrac{5}{2}) \rightarrow {^3\!}P_{2}\, (F=\nicefrac{7}{2})$ hyperfine transition. The horizontal lines indicate the time-cut applied for the projection onto the frequency axis shown in (b).
 The data are fitted using three Gaussian distributions representing the resonance signal (labeled {\it res. signal}) and contributions from the two background bands within the time window (labeled {\it bg band 1} and {\it bg band 2}). They sit on an additional background contribution from scattered laser light described by a second order polynomial (labeled {\it laser bg}).
}
 \label{fig:overnight_contour}
\end{figure}

As the robust determination of the resonance position is of primary interest, the analysis is carried out by projecting an appropriate time slice onto the frequency axis, as illustrated in Fig.\,\ref{fig:overnight_contour}b for the time window marked by white lines in Fig.\,\ref{fig:overnight_contour}a. Gating on this short time interval is essential for weaker hyperfine transitions (other than the shown $^{3\!}S_{1}\, (F=\nicefrac{5}{2}) \rightarrow {^3\!}P_{2}\, (F=\nicefrac{7}{2})$ component), to obtain a reasonable signal-to-noise ratio.

\subsection{Background fit}
The count rate observed in the \textit{background window}, i.e., in the time intervals from 0 to \SI{4}{\micro\second} and from 6 to \SI{10}{\micro\second}, during which no ions reach the detection chamber, originates purely from scattered laser light and from the photomultiplier dark counts. The same background level is expected within the \textit{analysis window}. When projecting the events from the background window onto the frequency axis, a slight decrease in the count rate toward higher frequencies becomes visible, attributed to variations in laser power during the scan, as power optimization was always performed at the scan’s starting frequency. 
The background information from this window is therefore modeled by fitting a second-order polynomial $P(x, c_2, c_1, c_0)$ to the frequency dependence using standard $\chi^2$ minimization, yielding reduced $\chi^2$ values close to unity. For the subsequent resonance fits, this background function is kept fixed after being rescaled to account for the different exposure times in the analysis and background windows.\ \\

\subsection{Signal fit model}
A triple Gaussian model has been found to reliably describe the resonance structures obtained by projecting the count-rate data onto the frequency axis.
The explicit model $M_j$ for the projection of a single run $j$ is:
\begin{eqnarray}
    M_j(x, \mathcal{G}_{1,j}, \mathcal{G}_{2,j}, \mathcal{G}_{3,j},P_j) &=& \sum_{i=1}^3 \mathcal{G}_{i,j}(x, \mu_{i,j}, \sigma_{i,j}, A_{i,j}) \nonumber \\
    & & + P_j(x, c_2, c_1, c_0) \hspace{1cm}
\end{eqnarray}

Here, $\mathcal{G}_{i,j}$ are the 3 Gaussian functions and $P_j$ is the background polynomial for run $j$, respectively.
$x$ is the scanning frequency.
Furthermore, the parameters $\mu_2$ and $\mu_3$ are substituted by their respective distances to $\mu_{1,j}$, and the amplitudes $A_{2,j}$ and $A_{3,j}$ are re-parameterized as their relative strength $f_{12}$ and $f_{13}$ with respect to $A_{1,j}$
\begin{align*}
	\mu_{2,j}   & = \mu_{1,j} - \Delta \mu_{1 2}      \\
	\mu_{3,j}   & = \mu_{1,j} -  \Delta \mu_{1 3}       \\
	A_{2,j} & = A_{1,j} \cdot f_{1 2} \\
	A_{3,j} & = A_{1,j} \cdot f_{1 3}.
\end{align*}
This parameterization enables a combined fit of several runs $j$ with shared parameters $\Delta \mu_{1 2}, \Delta \mu_{1 3}, f_{1 2}$ and $f_{1 3}$.
Likewise, the parameters of the Gaussian widths $\sigma_{i,j} = \sigma_{i}$ are shared between runs.
The corresponding $\chi^2$ function for such a combined fit of several runs is defined as 
\begin{equation}
    \label{eq:chi_2}
    \chi^2(\vec{p}) = \sum_{j=1}^{N} \sum_i^{N_{j}} \left(\frac{{M}({x_{j,i}}, \vec{p}) - {y_{j,i}}}{\mathrm{d}{y_{j,i}}}\right)^2  
\end{equation}
with the following symbols:
\vspace{0.5cm}\\
\begin{tabular}{ll}
	${x_{j,i}}$             & frequency step $i$ of run $j$                      \\
	$\vec{p}$               & parameter vector of all fit parameters             \\
	${M}$                   & fit model evaluated for scanning steps of run $j$  \\
	${y_{j,i}}$             & counts of file $j$                                 \\
    $\mathrm{d}{y_{j,i}}$   & error of counts (from Poisson statistics)          \\
    $N_{j}$                 & number of steps for run $j$                        \\
    $N$                     & number of runs
\end{tabular}
\vspace{0.5cm}\\

Fitting is done with {\em lmfit}~\cite{Newville2014}, which simplifies the re-parameterization and internally uses minimization routines from {\em scipy.optimize}~\cite{SciPy2020}.
The result for the reference data is shown in Fig.\,\ref{fig:overnight_contour}b. The individual three Gaussian peaks are indicated as well as their sum, which, in combination with the frequency-dependent background represents a reasonable description of the experimental data.

\subsection{Analysis Window Size} 
\begin{figure}[htp]
 \centering
 \includegraphics[width=\columnwidth]{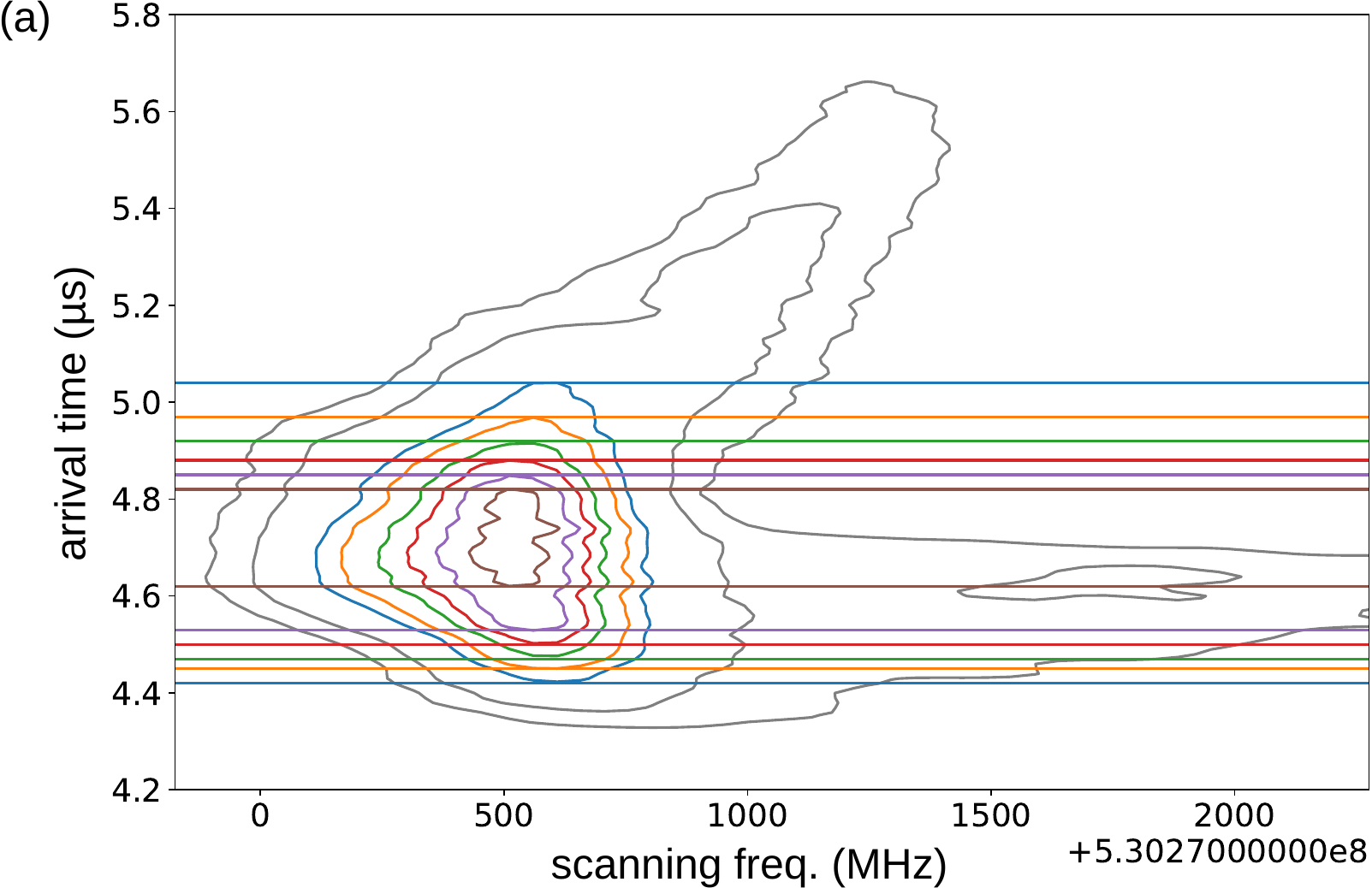} 
 \vspace{-1mm} 
 \includegraphics[width=\columnwidth]{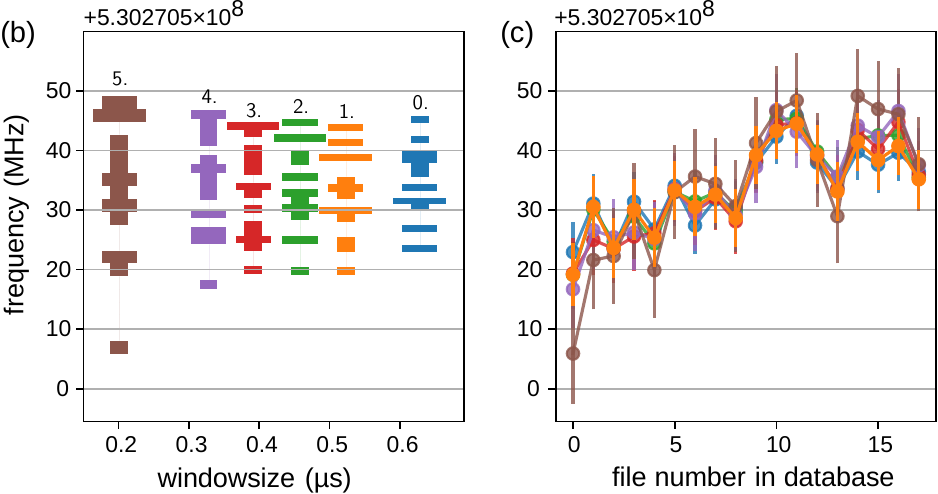}
 \caption{(a) Contour plot of the $^{3\!}S_{1}\, (F=\nicefrac{5}{2}) \rightarrow {^3\!}P_{2}\, (F=\nicefrac{7}{2})$ transition obtained from the sum of 18 consecutive measurement runs lasting approximately 30\,min each. To obtain the contour lines, the signal was blurred using a Gaussian filter with a (\SI{50}{\MHz}, \SI{20}{\nano\second}) kernel before plotting and extraction of borders. The horizontal lines indicate the various analysis windows (see text) resulting from the lowermost and uppermost point of the corresponding contour line. (b) Extracted frequencies for the transition in the 18 individual runs for different analysis windows corresponding to the 6 innermost contour lines in (a). The horizontal bars indicate the number of fits falling in the same bin (up to 3), while their vertical extent corresponds to the bin size in MHz. (c) Frequencies plotted as a function of the run number. A drift of the resonance frequency is observed for all time windows. 
 }
 \label{fig:TimeSliceDetermination}
\end{figure}

Since arrival time, velocity, and thus resonance frequency can be correlated, the choice of the analysis time slice may influence the extracted resonance frequency. To investigate this effect and to determine an optimal window that minimizes the amount of background while preserving the signal, the following procedure was applied: 18 consecutive runs of the $^{3\!}S_{1}\, (F=\nicefrac{5}{2}) \rightarrow {^3\!}P_{2}\, (F=\nicefrac{7}{2})$ transition were analyzed using six different window sizes, each defined by the uppermost and lowermost points of selected contour lines of the central resonance structure as shown in Fig.\,\ref{fig:TimeSliceDetermination}a. The two outermost contour lines were excluded, as they contain large portions of the tail from late-arriving ions.\\
For each window, the fitted center frequencies of the 18 runs are shown in Fig.\,\ref{fig:TimeSliceDetermination}b. The values are grouped into 20 bin histograms with the width of the horizontal bars indicating the number of runs contributing to the respective frequency bins. No systematic trend with window size is observed, though the frequency spread slightly decreases for larger windows. In Fig.\,\ref{fig:TimeSliceDetermination}c, the center frequencies are plotted versus run number, revealing a common drift across all windows, likely due to a fluctuation of the effective source potential as discussed in Appendix~\ref{sec:sys_stability_of_electroncurrent}.\\
The window labeled 1 in Fig.\,\ref{fig:TimeSliceDetermination}b, ranging from \SI{4.45}{\micro\second} to \SI{4.97}{\micro\second}, was chosen for all measurements because it yields the smallest frequency spread except for window 0, while providing a better reduced $\chi^2$ value of the fit compared to the latter. 

\section{Determination of Resonance Frequencies}
\label{sec:resonance_freq_calculation}
\begin{figure*}[htbp]
 \centering
 \includegraphics[width=\textwidth]{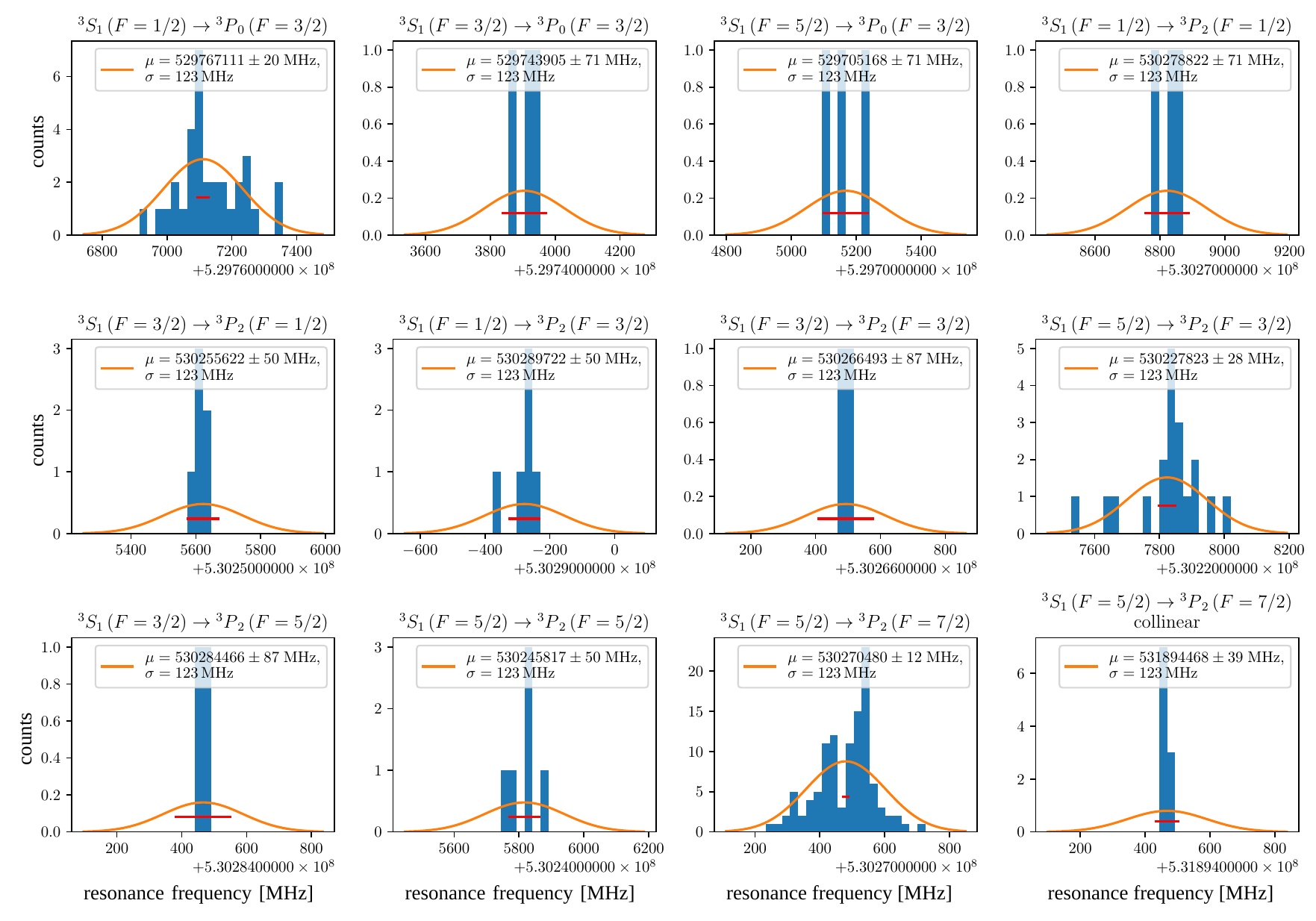}
 \caption[Histograms of transitions]{Histograms of the measured resonance frequencies for each transition as given by the wavemeter for the dye laser frequencies before second-harmonic generation. The histograms are fitted with Gaussian distributions (orange lines) using the unbinned likelihood method with a shared width $\sigma$. The uncertainties of the fitted central frequencies $\mu$ are indicated by red bars.
    }
    \label{fig:Resonance_histograms}
\end{figure*}

The in-principle constant parameters $\Delta \mu_{1 2}$, $\Delta \mu_{1 3}, f_{1 2}$, $f_{1 3}$ and $\sigma_{1,2,3}$ obtained from the 18 high-statistic runs were used to construct constraints $\mathcal{L}$ for the individual fits of the other, weaker transitions. 
These are called Lagrange multipliers 
\begin{equation}
    \mathcal{L} = \sum_{\substack{z=\left\{\Delta \mu_{1 2}, \Delta \mu_{1 3}, \right.\\ \left.f_{1 2}, f_{1 3},\sigma_{1,2,3}\right\}}} \left(\frac{z-z_P}{\Delta z_P}\right)^2\,.
\end{equation}
with the pull parameters $z_P$ and uncertainties $\Delta z_P$.
The different runs are then fitted independently with the $\chi^2$ function
\begin{equation}
    \chi^2(\vec{p}) = \sum_i^{N_\text{steps}} \left(\frac{M({x_i}, \vec{p}) - {y_i}}{\mathrm{d}y_i} \right)^2 + \mathcal{L} \; .
\end{equation}
Following this procedure, the center frequencies obtained in all individual runs are sorted into histograms belonging to the individual HFS-transitions observed during the campaign as depicted in Fig.\,\ref{fig:Resonance_histograms}.
The final resonance frequencies $\mu_t$ are then determined from unbinned likelihood fits for each histogram, with the unbinned likelihood function being
\begin{align}
    \text{ULL} =  \sum_t^T \sum_j^{N_t} \log \mathcal{G}_t(\mu_j, \mu_t, \sigma)
,\end{align}
with $T=12$ being the number of measured transitions, $\mu_j$ the center frequency of run $j$, $N_t$ the number of runs for the respective transition $t$, $\mu_t$ and $\sigma$ the fitting parameters to be optimized and $\mathcal{G}_t$ being a normalized Gaussian function evaluated at $\mu_j$. 
This method omits the fit-position uncertainties of individual runs. Because the run-center uncertainties $\mu_j$ are typically only a few MHz, systematic variations dominate, as seen in Fig.\,\ref{fig:Resonance_histograms}. In the unbinned likelihood fits, the width parameter $\sigma$ is shared, since rarely measured transitions should have the same spread as the frequently measured ones. The fitted center values $\mu_t$ and their uncertainties then provide the final (not yet Doppler-corrected) resonance frequencies for each transition.

Besides the statistical contributions, the uncertainties from the likelihood fit incorporate the influence of systematic effects which fluctuated during the experimental campaign, details are discussed in Appendix~\ref{sec:systematics}. 
An independent systematic uncertainty of $\Delta\nu \leq 20\;$MHz comes from the frequency measurement with the wavemeter (Appendix~\ref{sec:wavemeter}). This uncertainty is also included in the final results presented in Table\,\ref{tab:resonance_freqs}. \\
\begin{table*}[htbp]
    \centering
    \caption{Resonance frequencies in the laboratory system and in the ion rest frame. All values are in MHz.}
    \label{tab:resonance_freqs}
    \begin{tabular}{lll}
        \toprule
        Transition  & Resonance frequency & Resonance frequency \\ 
	                  & (laboratory system) & (ion rest frame) \\
        \midrule
        $^{3\!}S_{1} (F=\nicefrac{5}{2}) \rightarrow {^3\!}P_{0} (F=\nicefrac{3}{2})$  & \SI{1059410336 (143)}{} & \SI{1061031352 (189)}{} \\
        $^{3\!}S_{1} (F=\nicefrac{3}{2}) \rightarrow {^3\!}P_{0} (F=\nicefrac{3}{2})$  & \SI{1059487810 (143)}{} & \SI{1061108945 (189)}{} \\
        $^{3\!}S_{1} (F=\nicefrac{1}{2}) \rightarrow {^3\!}P_{0} (F=\nicefrac{3}{2})$  & \SI{1059534223 (46) }{} & \SI{1061155429 (132) }{} \\
        $^{3\!}S_{1} (F=\nicefrac{5}{2}) \rightarrow {^3\!}P_{2} (F=\nicefrac{3}{2})$  & \SI{1060455647 (60) }{} & \SI{1062078263 (138) }{} \\
        $^{3\!}S_{1} (F=\nicefrac{5}{2}) \rightarrow {^3\!}P_{2} (F=\nicefrac{5}{2})$  & \SI{1060491634 (102)}{} & \SI{1062114304 (161)}{} \\
        $^{3\!}S_{1} (F=\nicefrac{3}{2}) \rightarrow {^3\!}P_{2} (F=\nicefrac{1}{2})$  & \SI{1060511245 (102)}{} & \SI{1062133946 (161)}{} \\
        $^{3\!}S_{1} (F=\nicefrac{3}{2}) \rightarrow {^3\!}P_{2} (F=\nicefrac{3}{2})$  & \SI{1060532986 (175)}{} & \SI{1062155720 (215)}{} \\
        $^{3\!}S_{1} (F=\nicefrac{5}{2}) \rightarrow {^3\!}P_{2} (F=\nicefrac{7}{2})$  & \SI{1060540960 (31) }{} & \SI{1062163706 (124) }{} \\
        $^{3\!}S_{1} (F=\nicefrac{1}{2}) \rightarrow {^3\!}P_{2} (F=\nicefrac{1}{2})$  & \SI{1060557645 (143)}{} & \SI{1062180416 (190)}{} \\
        $^{3\!}S_{1} (F=\nicefrac{3}{2}) \rightarrow {^3\!}P_{2} (F=\nicefrac{5}{2})$  & \SI{1060568932 (175)}{} & \SI{1062191721 (215)}{} \\
        $^{3\!}S_{1} (F=\nicefrac{1}{2}) \rightarrow {^3\!}P_{2} (F=\nicefrac{3}{2})$  & \SI{1060579443 (102)}{} & \SI{1062202248 (161)}{} \\
        $^{3\!}S_{1} (F=\nicefrac{5}{2}) \rightarrow {^3\!}P_{2} (F=\nicefrac{7}{2})$ (collinear) & \SI{1063788935 (80)}{} & \\    
    \bottomrule
    \end{tabular}
\end{table*}
The rest-frame frequency $\nu_0$  of a transition can be calculated from measurements of the laboratory frequencies in collinear $\nu_c$ and anti-collinear $\nu_a$ geometry using
\begin{equation}
    \nu_0 = \sqrt{\nu_a \nu_c} \; .
\end{equation} 
The uncertainty related to deviations from a perfectly parallel geometry is discussed in Appendix~\ref{sec:systematics_angle}. The Doppler factor $D$\replaced{, describing the frequency shift}{} is then expressed as
\begin{equation}
    \label{eq:Doppler_factor}
    D  := \frac{1}{\gamma(1-\beta \cos\theta)} = \frac{\nu_0}{\nu_a} = \frac{\sqrt{\nu_a \, \nu_c} }{\nu_a} = \sqrt{\frac{\nu_c}{\nu_a}}
\end{equation}
and is used to transform the anti-collinear resonance frequency $\nu_{a,t}$ of a transition $t$ to the center of mass frame using
\begin{equation}
    \nu_{0,t} = D\cdot \nu_{a,t} \,.
\end{equation} 
A measurement of both frequencies, $\nu_a$ and $\nu_c$, has been performed for the strong $^{3\!}S_{1} (F=\nicefrac{5}{2}) \rightarrow {^3\!}P_{2} (F=\nicefrac{7}{2})$ transition. As the collinear measurement required the temporary installation and adjustment of an additional mirror in the beamline above the detection system, it was only performed once during the week of measurements in three consecutive runs. 
Since the source potentials were not changed during the measurement campaign, the same Doppler factor is applied to all measured transitions. To take the observed fluctuations over the week of measurements into account in the calculation of the uncertainty of $D$, we conservatively use the fitted width $\sigma$ as the uncertainty of the collinear frequency. We then obtain the Doppler factor

\begin{equation}
    D = \SI{1.00153011 \pm 0.00000012}{} \; .
\end{equation}

The resulting center-of-mass frequencies are listed in Table\,\ref{tab:resonance_freqs}.

\section{Calculation of Hyperfine Parameters}

The hyperfine structure in the $^3\!P$ level of He-like B is disturbed by level mixing. Without hyperfine structure, only the fine-structure mixing between the $^1\!P_1$ and the $^3\!P_1$ level would have to be taken into account, but the hyperfine-induced level mixing affects all fine-structure components since they all contain levels with identical total angular momentum $F$. Their smaller energy separation, compared to the $^1\!P$\,--\,${}^3\!P$ splitting, enhances this effect. This has to be taken into account in the analysis. 

The commonly used magnetic hyperfine constant 
\begin{align}
    A_\mathrm{hfs} &= \frac{\operator{J J}{J J}{\mathcal{T}_\mathrm{e}}{1} \operator{I I}{I I}{\mathcal{T}_\mathrm{n}}{1}}{I J} \nonumber \\
    &= \begin{pmatrix} J & 1 & J \\ J & 0 & -J \end{pmatrix} 
    \cdot \frac{\irreducableoperator{J}{J}{\mathcal{T}_\mathrm{e}}{1}\:g_I \, \mu_N}{J} \nonumber \\
    &= \frac{\irreducableoperator{J}{J}{\mathcal{T}_\mathrm{e}}{1} \: g_I \, \mu_N}{\sqrt{(2 J + 1)\,J\,(J + 1)}}
\label{eq:A_hfs}
\end{align}
quantifies the interaction of the nuclear magnetic dipole moment $\mu_I = \operator{I I}{I I}{\mathcal{T}_\mathrm{n}}{1} = g_I \, I \, \mu_N $ and the magnetic field generated by the electrons of a certain fine-structure state with total orbital angular momentum $J$ at the position of the nucleus. This magnetic field is calculated by the reduced matrix element of the electronic magnetic dipole operator $\irreducableoperator{J}{J}{\mathcal{T}_\mathrm{e}}{1}$. If the energy spacing between the involved fine structure states is sufficiently large and the magnitude of the magnetic field generated by the electrons is small, the magnetic hyperfine structure energy can be calculated by the formula
\begin{align}
    E_\mathrm{hfs} = \frac{A_\mathrm{hfs}}{2} \, \left[ F \, (F + 1) - I \, (I + 1) - J \, (J + 1)\right].
\label{eq:E_hfs}
\end{align}

\begin{table}[ht]
\caption{Values of the matrix elements $\langle\gamma J\|\mathcal{T}^{(1)}_\mathrm{e}\| \gamma^{\prime} J^{\prime}\rangle \quad \equiv$ $\langle(2 S+1) J\|\mathcal{T}^{(1)}_\mathrm{e}\|\left(2 S^{\prime}+1\right) J^{\prime}\rangle$ for the $1s2p\,{}^{3\!}P_{0,1,2}$ and ${}^{1\!}P_1$ states in ${ }^{11}\mathrm{B}^{3+}$ taken from~\cite{Johnson1997}. All values are given in units of $\nicefrac{e}{\left( 4 \pi \varepsilon_0 c a_0^2 \right)}= 1715.256\,\mathrm{T}$, where $a_0$ is the Bohr radius.}
\label{tab:nondiagonal_elements}
\begin{tabular}{lrrrr}
\hline $\mathcal{T}^{(1)}_\mathrm{e}$ & \multicolumn{1}{c}{$\| 30\rangle$} & $\| 31\rangle$ & $\| 32\rangle$ & \multicolumn{1}{c}{$\| 11\rangle$} \\
\hline$\langle 30 \|$ & 0 & 1.6539 & 0 & --1.1828 \\
$\langle 31 \|$ & 1.6539 & 1.5472 & 1.8712 & --2.1101 \\
$\langle 32 \|$ & 0 & 1.8712 & 3.3485 & 2.7069 \\
$\langle 11 \|$ & --1.1828 & --2.1101 &  2.7069 & 0.032101 \\
\hline
\end{tabular}
\end{table}

However, in triplet states, the parallel alignment of the electron spins gives rise to a particularly strong magnetic field and the separation between the fine structure levels is comparatively small. As a result, off-diagonal matrix elements of the magnetic dipole hyperfine operator are significantly enhanced and lead to a hyperfine-induced fine-structure mixing between substates of the $2\,^{1,3\!}P$ states. Therefore, the complete Hamiltonian $H=H_0+H_{\mathrm{hfs}}$ with the fine-structure Hamiltonian $H_0$ and the hyperfine interaction $H_\mathrm{hfs}$ needs to be diagonalized simultaneously to evaluate the energy eigenvalues of the coupled states with total angular momentum $ F $. This leads to a system of eigenvalue equations~\cite{Johnson1997}
\begin{eqnarray}
E_{\gamma J}^F C_{\gamma J}^F &=& E_{\gamma J} C_{\gamma J}^F +  \\\nonumber
&& \sum_{\substack{\gamma^{\prime}, J^{\prime} \\ J^{\prime}=F-I}}^{F+I}\langle\alpha I, \gamma J ; F m| H_{\mathrm{hfs}}\left|\alpha I, \gamma^{\prime} J^{\prime} ; F m\right\rangle C_{\gamma^{\prime} J^{\prime}}^F
\label{eq:full_HFS}
\end{eqnarray}
that must be solved. Here, $E_{\gamma J}$ are the eigenvalues of the $F$-independent Hamiltonian $H_0$ and $C_{\gamma J}^F$ are the components of the eigenvectors of $H$.
In contrast to Eq.\,(\ref{eq:E_hfs}), off-diagonal terms are considered. In the coupled-state basis this leads to the following hyperfine structure energy\\
\begin{widetext}
    \begin{align}
        E_\mathrm{hfs} &= (-1)^{I + J + F} \, \left\{\begin{matrix} I & J & F \\ J^\prime & I & 1 \end{matrix}\right\} \, \times \irreducableoperator{\gamma J}{\gamma^\prime J^\prime}{\mathcal{T}_\mathrm{e}}{1}\irreducableoperator{I}{I}{\mathcal{T}_\mathrm{n}}{1}
        \nonumber \\
        \nonumber \\
        &= (-1)^{I + J + F} \, \left\{\begin{matrix} I & J & F \\ J^\prime & I & 1 \end{matrix}\right\} \, \times \irreducableoperator{\gamma J}{\gamma^\prime J^\prime}{\mathcal{T}_\mathrm{e}}{1} \sqrt{(2 I + 1)\,I\,(I + 1)} \, g_I \, \mu_N.
        \label{eq:E_hfs_product}
    \end{align}
\end{widetext}
Note that for vanishing off-diagonal matrix elements Eq.\,(\ref{eq:E_hfs_product}) is identical to Eq.\,(\ref{eq:E_hfs}).\\
Since we were not able to measure all hyperfine components of the manifold, but only those of the $^3\!P_2$ and $^3\!P_0$ levels, we require theoretical input to include the contributions of the other two levels and extract the undisturbed $A_\mathrm{hfs}(^3\!P_2)$ and the energies of the fine-structure levels. Thus, we use the nondiagonal matrix elements of the magnetic dipole operator $\mathcal{T}^{(1)}_\mathrm{e}$ of the $1s2p\,^{3\!}{P}_{0,1,2}$ and $1s2p\,^{1\!}P_1$ states in He-like systems from~\cite{Johnson1997}, which were evaluated using relativistic conﬁguration-interaction wave functions that account for both Coulomb and Breit interactions, and also include QED terms. The values for $^{11}\mathrm{B}^{3+}$ used in this work are listed in Table~\ref{tab:nondiagonal_elements}. 
The nuclear magnetic moment $\mu_I (^{11}\mathrm{B})=2.688378(1)\,\mu_N $ has been taken from~\cite{Jackowski2009}. To account for a possible hyperfine structure anomaly and a deviation from the theoretical value due to lacking higher-order QED contributions, we use the diagonal matrix elements as adjustable parameters for our fitting routine, while the off-diagonal terms are fixed to the theoretical values. Other parameters of our fit function are the fine-structure transition frequencies and the electric quadrupole hyperfine structure constant $B(^3\!P_2)$. The electric quadrupole constants of the levels that we have not addressed, were fixed to $B({}^{1,3\!}P_1) =\: $\SI{0}{MHz}. 
The influence of the ${}^{3\!}P_1$ and ${}^{1\!}P_1$ levels on the hyperfine structure of the investigated $^{3\!}P_{0,2}$ levels was accounted for by including the energies of all relevant fine-structure levels in the diagonalization performed during the fit. While the energies of the $^{3\!}P_{0,2}$ levels were treated as free parameters, the energy of the fine-structure states $^{1,3\!}P_1$, which have not been probed, and the matrix elements of the ${}^{1,3\!}P_1$ hyperfine levels were fixed to the literature values provided in~\cite{Dinneen1991, Yerokhin.2010} and~\cite{Johnson1997}, respectively.
Results for the free fitting parameters are summarized in Table~\ref{tab:fit_parameter}. The matrix element $\irreducableoperator{31}{31}{\mathcal{T}_\mathrm{e}}{1}$ was not fitted directly, but calculated scaling $\irreducableoperator{32}{32}{\mathcal{T}_\mathrm{e}}{1}$ by the ratio of the theoretical values listed in Table\,\ref{tab:nondiagonal_elements}. 
\begin{table}[bh]
\caption{Fit parameters and their values resulting from the analysis. }
\label{tab:fit_parameter}
\begin{tabular}{cr}
\toprule
  Fit parameter  &  \multicolumn{1}{c}{value}  \\
\midrule
    $\nu({}^{3\!}S_1\rightarrow {}^{3\!}P_0)$  & \SI{1061074196+-103}{MHz} \\
    $\nu({}^{3\!}S_1 \rightarrow {}^{3\!}P_2)$  & \SI{1062167247+-68}{\phantom{0}\MHz}\\
    $B({}^{3\!}P_2)$  & \SI{-28+-124}{\MHz} \\
    $A({}^{3\!}S_1)$  & \SI{31001+-37}{\phantom{0}\MHz}\\
    $\irreducableoperator{31}{31}{\mathcal{T}_\mathrm{e}}{1}$  & $1.547(3) \cdot e/(4 \pi \varepsilon_0 c a_0^2)$\\
    $\irreducableoperator{32}{32}{\mathcal{T}_\mathrm{e}}{1}$  & $3.348(6)\cdot e/(4 \pi \varepsilon_0 c a_0^2)$\\
\bottomrule
\end{tabular}
\end{table}
\section{Discussion and Interpretation}
Previous experimental results for the hyperfine splitting in the triplet states of He-like $^{11}{\rm B}^{3+}$ have been presented in~\cite{Dinneen1991}. In the analysis, Dinneen \textit{et al}.\ used radial integrals of the relativistic single electron wavefunctions calculated by~\cite{Schwartz1955, Schwartz1955} and the theory of the hyperfine structure for triplet states in He-like ions presented in~\cite{Lurio1962}. They further decompose the reduced magnetic dipole matrix element of the coupled electronic states into the sum of matrix elements of the single electron wave functions. As a result, they obtain the single electron magnetic dipole constants $a_j$ with $j$ being the total angular momentum of the single electron wave function. In this decomposition the off-diagonal matrix element connecting the $p_{1/2}$ and $p_{3/2}$ components carries the relativistic factor $ \xi $, defined as the ratio of the off-diagonal to the diagonal radial dipole integral~\cite{Schwartz1955}. In the expression of the magnetic hyperfine structure constant $A_\mathrm{hfs} (^3P_1)$ the relativistic factor $ \xi $ accounts for the relativistic correction due to the admixture between $p_{1/2}$ and $p_{3/2}$. In the non-relativistic limit this correction factor tends to unity.  Finally, following the procedure in~\cite{Lurio1962} they calculated the hyperfine constants $A_\mathrm{hfs}$ that can be compared to our results. At least during this conversion into commonly used hyperfine structure constants, however, the $ \xi $-dependent cross term was not considered correctly in their analysis. While $A_\mathrm{hfs} (^3S_{1})$ and $A_\mathrm{hfs} (^3P_{2})$ do not depend on $\xi$, and are therefore reproducible from their single-electron magnetic dipole constants, this does not hold for $A_\mathrm{hfs} (^3P_{1})$. Here, the presented value is only recovered when setting $ \xi $ to the unphysical value zero. Their results are compiled together with the earlier calculations by~\cite{Aashamar1977} and our results in Table~\ref{tab:a_factors}. \\
\begin{table}[tb]
 \caption{Comparison of magnetic hyperfine structure constants. All units are given in MHz. 1-$\sigma$ uncertainties are quoted in parantheses.}
 \label{tab:a_factors}
 \begin{tabular}{ccccc}
  \toprule
  hyperfine & this work & Dinneen & Aashamar &  Johnson \\
  constant && \cite{Dinneen1991} & \cite{Aashamar1977} & \cite{Johnson1997}\\
  \midrule
  $A({{}^3\!}S_1)$  & \SI{31001+-35}{} & \SI{31084+-40}{} & \SI{30930}{} & --\\
  $A({}^{3\!}P_1)$  & \SI{14800+-26}{} & \SI{14788+-40}{} & \SI{14648}{} & \SI{14801}{} \\
  $A({}^{3\!}P_2)$  & \SI{14324+-25}{} & \SI{14366+-20}{} & \SI{14302}{} & \SI{14326}{} \\
  \bottomrule
 \end{tabular}
\end{table}
\begin{table}[htbp]
 \centering
 \caption[Energies of fine structure levels with respect to the singlet ground state $^3\!S_1$ in MHz.]
 {Energies of fine structure levels w.r.t. the singlet ground state $^3\!S_1$ in MHz. 1-$\sigma$ uncertainties are given in parantheses, and adapted for level energies presented in~\cite{Dinneen1991} accordingly.}
 \label{tab:final_finestructures}
 \begin{tabular}{cccc}
 \toprule
Level & Fit (this work) & Dinneen~\cite{Dinneen1991} & Yerokhin~\cite{Yerokhin.2010} \\
\midrule
 ${^1{\!}}P_1$ & --                     & --                     & \SI{1691863948+-420}{} \\
 ${}^{3\!}P_2$ & \SI{1062167248+-69}{}  & \SI{1062167197+-90}{} & \SI{1062167317+-420}{} \\
 ${}^{3\!}P_0$ & \SI{1061074196+-102}{}  & \SI{1061074244+-130}{} & \SI{1061074274+-420}{} \\
 ${}^{3\!}P_1$ & --                     & \SI{1060588490+-130}{} & \SI{1060588640+-420}{} \\
\bottomrule 
\end{tabular}
\end{table}

Overall, there is reasonable agreement between our results and those of~\cite{Dinneen1991} within the combined uncertainties. Our hyperfine structure constant $A({}^{3\!}S_1)$ tends to have a slightly lower value compared to the previous measurement. Excellent agreement is observed between experimental results and hyperfine-structure predictions of~\cite{Johnson1997}, when evaluated with the magnetic moment reported in~\cite{Jackowski2009}. 
\\
As shown in Table~\ref{tab:final_finestructures}, the resulting values for the energies of the ${}^{3\!}P_2$ and ${}^{3\!}P_0$ states are in good agreement with the literature values.
Taking the more recent calculations of~\cite{Johnson1997} into consideration, our extracted center-of-gravity (cg) frequencies are not affected as strong by an incomplete decomposition of the hyperfine structure as they were in the analysis of~\cite{Dinneen1991}. Additionally, we used the electric quadrupole hyperfine constant $B({}^{3\!}P_2)$ as free parameter instead of fixing it to its theoretical prediction. The validity of our procedure has been also confirmed by an accurate determination of the cg frequencies in $^{13}$C~\cite{mueller2025} and the corresponding agreement of the splitting isotope shift in the $^{12,13,14}$C$^{4+}$ ions~\cite{Mueller.2026}.   

\section{Conclusions}
We have measured the $1s2s\,{}^{3\!}S_1 \rightarrow 1s2p\,{}^{3\!}P_{0,2}$ transitions in He-like $^{11}$B$^{3+}$ and obtained slightly improved accuracies compared to previous measurements in~\cite{Dinneen1991}. The production process of He-like B$^{3+}$ in an EBIT and the population of the $1s2s\,{}^{3\!}S_1$ level using trimethylborate in a MIVOC setup was demonstrated and an analysis procedure for the complex line profile of the bunches extracted from the SPARC-EBIT was established that can be used in further investigations at HITRAP. However, production rates and signal levels have been rather low and did not allow for a measurement of the less-abundant isotope $^{10}$B. Therefore, tests with other evaporable boron composites are ongoing at the COALA beamline with promising progress. There, the beamline and the available electron beam ion source EBIS-A provide better conditions for high-precision measurements~\cite{Imgram.2023b}. Moreover, we will overcome the current experimental limitations by quasi-simultaneous collinear and anticollinear laser spectroscopy referenced to a frequency comb as has been demonstrated in~\cite{Imgram2023,mueller2025,Koenig2026}, where 2-MHz precision has been achieved for He-like carbon ions.\\
Additionally, we want to stress the point that the most recent theory of the hyperfine structure in He-like boron does not include higher-order QED corrections, which are nowadays considered. Modern NRQED calculations up to the order of $ m\alpha^6$ were conducted for Li$^+$ and Be$^{2+}$~\cite{QiZhang2020,QiZhang2023}, but are missing for B$^{3+}$. In the case of Li$^+$, it has been demonstrated that the current theoretical accuracy is sufficient to extract the Zemach radius by comparing the experimental value of the $^3\!S$ hyperfine splitting with the theoretical prediction that omit nuclear corrections based on the Zemach radius~\cite{QiZhang2020,GuanChen2020}. 
Since the sensitivity parameter to the Zemach radius increases with the proton number $Z$~\cite{QiZhang2020}, an improved experimental determination of the hyperfine structure in B$^{3+}$ in combination with more precise theoretical predictions will enable the extraction of the Zemach radius for boron isotopes as well. As demonstrated for $^6$Li$^+$~\cite{QiZhang2020, GuanChen2020}, this parameter is sensitive to the nuclear structure. Together with the root-mean-square charge radius, the Zemach radius can provide complementary information about the nuclei, especially the proton-halo candidate $^8$B. 
\section*{Data availability statement}
Data sets generated during the current study are available from the corresponding author on reasonable request.\\

\begin{acknowledgments}
We acknowledge support by the BMBF under contract numbers 05P19RDFAA, 05P21RDFA1,05P24RD5 and 05P19PMFA1 and by the Deutsche Forschungsgemeinschaft (DFG, German Research Foundation) -- Project-Id 279384907—SFB 1245.
\end{acknowledgments}
\appendix
\section{Systematics}
\label{sec:systematics}
The uncertainty of the transition frequencies in the laboratory system including systematic effects has been estimated in Section~\ref{sec:resonance_freq_calculation} based on the spread of the measured resonance frequencies throughout the campaign since the dominant systematics varied on a day-to-day basis. The following main systematic effects are expected to contribute to the observed spread:
\begin{enumerate}
    \item interaction angle between laser and ion beam \label{itm:angle}
	\item acceleration voltage stability \label{itm:voltage}
	\item electron beam current stability \label{itm:current}
	\item laser frequency stability  \label{itm:frequency}
\end{enumerate}
While the effects \ref{itm:angle} to \ref{itm:current} affect the Doppler correction,  effect \ref{itm:frequency} is only dependent on the accuracy of the wavemeter used for frequency monitoring and stabilization.

\subsection{Interaction Angle Between Laser and Ion Beam}
\label{sec:systematics_angle}
From Eq.\,(\ref{eq:Doppler-Shift}) we can deduce that a deviation $\Delta\theta$ from a perfectly (anti-)parallel superposition of laser and ion beams will cause a shift in the observed frequency given by 
\begin{equation}
    \Delta\nu_{a,c} = \pm \frac{\nu_{a,c}\beta(1 - \cos \Delta\theta)}{1 \pm \beta \cos\Delta\theta} \; .
\end{equation}
The maximum deviation of the ion and laser beams from the vertical axis of the detection chamber shown in Fig.\,\ref{fig:spectrap_detector} is defined by the \SI{10}{\milli\meter} diameter openings in the apertures mounted above and below the interaction region at a distance of \SI{636}{\milli\meter}. In addition, measurements with a camera mounted on top of the setup were used to make sure that the laser beam exited the interaction region centrally through the upper aperture such that only an off-centered entry through the lower aperture could contribute to an angular deviation from the vertical. Similarly for the ion beam the adjustable iris aperture on top together with an MCP detector below the setup was used to make sure the ions would enter the interaction region centered. Taking into account the observed diameter of the ion beam of \SI{4}{\milli\meter} and the diameter of the laser beam of \SI{3.2}{\milli\meter}, the maximum possible angular deviation between both is then given by
\begin{equation}
    \Delta\theta = \arctan\left(\frac{\SI{3}{\milli\meter}}{\SI{636}{\milli\meter}}\right) + \arctan\left(\frac{\SI{3.4}{\milli\meter}}{\SI{636}{\milli\meter}}\right) \approx \SI{10}{\milli\radian}
\end{equation}
resulting in a maximum frequency shift of \SI{81}{MHz}.
\subsection{Stability of the Acceleration Voltage}
\label{sec:sys_stability_of_voltage}
The acceleration voltage of the ion source was generated by a FuG HCN35M-6500 power supply and was not changed during the measurement campaign. The stability of the acceleration voltage was checked using a portable voltage divider built from spare resistors of the KATRIN K35 divider~\cite{Thuemmler2007} that was calibrated beforehand against the precision G35 divider~\cite{Winzen2021}. The divider output was measured using a Keysight 34465A digital voltmeter. \\
After switching on the power supply and allowing the voltage to settle during a 20-minute interval, a maximum drift of \SI{0.4}{V} was observed during a 19 hour interval.
Using Eq.\,(\ref{eq:DifferentialDopplerShift}) this corresponds to a maximum frequency drift also of \SI{81}{MHz}. We note that the measured voltage stabilized at \SI{4010.6}{V} with a remaining ripple of 0.03~V after 14 hours of the test measurement, which indicates that the actual voltage drifts during the measurement campaign may have been significantly lower.\\
Before the test of the power supply, it was running for months within an air-conditioned container, only interrupted once during the measurement campaign, due to a blackout at the facility. 
The reproducibility of a preset voltage after a power loss was also tested. With little cool-down time between switching off and on, voltage changes of up to \SI{0.1}{V} were observed after such a short power cycle.
\subsection{Stability of the Electron Beam Current}
\label{sec:sys_stability_of_electroncurrent}
\begin{figure}[h]
 \centering
 \includegraphics[width=\columnwidth]{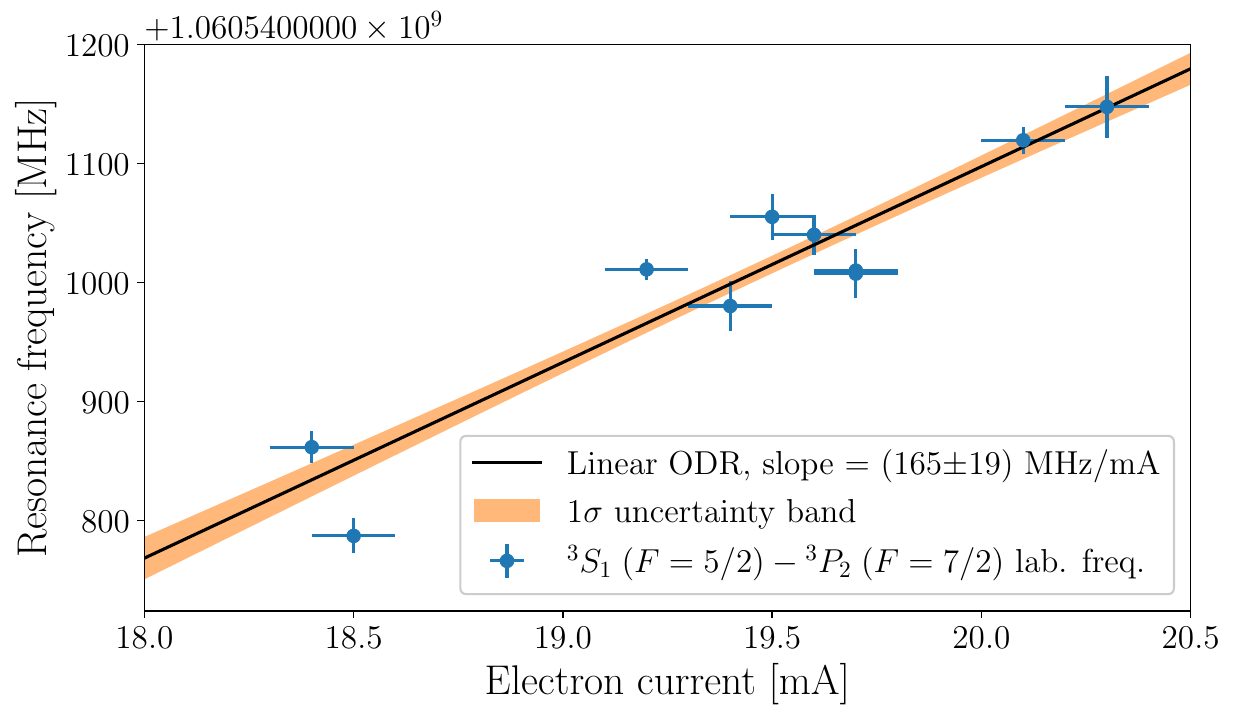}
 \caption{Measurements of the $^{3\!}S_{1} (F=\nicefrac{5}{2}) \rightarrow {^3\!}P_{2} (F=\nicefrac{7}{2})$ resonance frequency versus EBIT electron current. Error bars of the measured frequencies correspond to the statistical uncertainties only.
 A linear fit is applied to the data using the orthogonal-distance-regression method~\cite{Boggs1989}. 
 }
 \label{fig:systematics_ebit_current}
\end{figure}

The electric potential experienced by the ions in the EBIT is lower than the applied voltage due to the space-charge potential of the electron beam inside the source. According to~\cite{Penetrante1991}, the depth of the space-charge potential scales linearly with the EBIT's current. While care was taken to keep the electron current at \SI{19.5}{mA} manually, there was no automated control or monitoring of the value during the campaign. \\
Fig.\,\ref{fig:systematics_ebit_current} displays measurements of 
the  $^{3\!}S_{1}\, (F=\nicefrac{5}{2}) \rightarrow {^3\!}P_{2}\, (F=\nicefrac{7}{2})$ transition frequency in the laboratory system as a function of the electron beam current taken at different times during the campaign. 
An uncertainty of $\pm$\SI{0.1}{mA} is assigned to the current values, corresponding to the observed fluctuation of the current display of the power supply. 
A linear fit of the data-points using the orthogonal-distance-regression method~\cite{Boggs1989} results in a slope of $165 \pm 19\;$MHz/mA.
Given the range of currents occurring in the datasets used in the fit of $\approx \pm \SI{1}{mA}$, this is the dominant contribution to the width of the observed frequency distributions shown in Fig.\,\ref{fig:Resonance_histograms}.

\subsection{Measurement of the Laser Frequency}
\label{sec:wavemeter}
The laser frequency was continuously measured using a WS Ultimate-10 wavelength meter and actively stabilized to the set value with a digital PID feedback loop. The nominal precision of \SI{10}{MHz} (\SI{20}{MHz} after frequency doubling) defines the total-frequency uncertainty, which is estimated as the upper limit of the laser-frequency uncertainty since a direct comparison of the wavemeter reading with a frequency-comb measurement suggests a higher accuracy for the red to near infrared wavelength regime~\cite{Koenig2020}. We also note that this uncertainty will not contribute to the spread of the experimental results since the typically observed fluctuation of the offset on the timescale of the measurements (weeks) is below \SI{1}{MHz}. The wavemeter uncertainty is included in the error bars of the resonance frequencies in Table\,\ref{tab:resonance_freqs}. 
\bibliography{bibliography}
\end{document}